# Exploring thermal expansion of carbon-based nanosheets by machine-learning interatomic potentials

Bohayra Mortazavi [a,b,*], Ali Rajabpour[c], Xiaoying Zhuang[a,d,**], Timon Rabczuk[d] and Alexander V. Shapeev[e]

[a]Chair of Computational Science and Simulation Technology, Institute of Photonics, Department of Mathematics and Physics, Leibniz Universität Hannover, Appelstraße 11,30167 Hannover, Germany.
[b]Cluster of Excellence PhoenixD (Photonics, Optics, and Engineering–Innovation Across Disciplines), Gottfried Wilhelm Leibniz Universität Hannover, Hannover, Germany
[c]Mechanical Engineering Department, Imam Khomeini International University, Qazvin, Iran.
[d]College of Civil Engineering, Department of Geotechnical Engineering, Tongji University, 1239 Siping Road Shanghai, China.
[e]Skolkovo Institute of Science and Technology, Skolkovo Innovation Center, Nobel St. 3, Moscow 143026, Russia.

## Abstract

Examination of thermal expansion of two-dimensional (2D) nanomaterials is a challenging theoretical task with either ab-initio or classical molecular dynamics simulations. In this regard, while ab-initio molecular dynamics (AIMD) simulations offer extremely accurate predictions, but they are excessively demanding from computational point of view. On the other side, classical molecular dynamics simulations can be conducted with affordable computational costs, but without predictive accuracy needed to study novel materials and compositions. Herein, we explore the thermal expansion of several carbon-based nanosheets on the basis of machine-learning interatomic potentials (MLIPs). We show that passively trained MLIPs over inexpensive AIMD trajectories enable the examination of thermal expansion of complex nanomembranes over wide range of temperatures. Passively fitted MLIPs could also with outstanding accuracy reproduce the phonon dispersion relations predicted by density functional theory calculations. Our results highlight that the devised methodology on the basis of passively trained MLIPs is computationally efficient and versatile to accurately examine the thermal expansion of complex and novel materials and compositions using the molecular dynamics simulations.
Keywords: Thermal expansion; graphene; 2D materials; machine learning.

Corresponding authors: *bohayra.mortazavi@gmail.com; **zhuang@iop.uni-hannover.de

# 1. Introduction

Temperature is known to affect structural and intrinsic physical properties of materials. By heating a material, consisting atoms begin to move and vibrate more, leading to larger distance between themselves and as such normally materials expand with temperature. The change of a materials dimension with temperature is normally explained via the linear thermal expansion coefficient (TEC), which depending on the atomic interactions is typically a non-linear temperature dependent function. Shortly after the first report of graphene isolation [1], Mounet and Marzari [2] theoretically predicted that single-layer graphene shows an anomalous behavior and exhibits a negative TEC. Using the Raman spectroscopy technique, Moon *et al.* [3] measurements confirm the aforementioned theoretical prediction of negative TEC for graphene between 200–400 K. Later Raman measurements by Linas and coworkers [4] revealed a positive TEC for graphene for temperatures above 300 K. Nevertheless, in the both aforementioned experimental works the graphene was fully supported over a substrate. It is well-know that the substrate can strongly affect the properties of atomic-thick graphene. Tian *et al.* [5] employed Raman spectroscopy to measure the TEC of free-standing graphene monolayer, but they have yet required a theoretical correction for graphene-substrate mismatch in the evaluation of TEC. In a latest experimental work, Kano and coworkers [6] employed electron diffraction technique in order to evaluate the substrate and contamination effects on the TEC of suspended graphene.

As it is clear, the experimental measurement of graphene's TEC is a challenging task. Hence for the efficient examination of TEC of the other existing and novel two-dimensional (2D) materials, development of an accurate modeling approach is highly required. Classical molecular dynamics (MD) simulations and density functional theory (DFT) are the most common theoretical methods to evaluate various thermal properties, including the TEC. Nonetheless, for novel materials the empirical interatomic potentials are mostly unavailable, hindering the development of MD models. More importantly, MD estimations are directly dependent on the accuracy of employed force fields for describing the atomic interactions and therefore the accuracy of MD estimations cannot be unconditionally trusted. For example, the majority of available interatomic potentials for carbon atoms, fail to accurately reproduce the thermal conductivity of suspended graphene monolayer. First-principles DFT simulations on the other side can be conveniently employed to examine a wide range of properties of novel materials and compositions with outstanding level of accuracy and reproducibility. The computational cost of DFT simulations however scales fast with the number of atoms, and they are thus limited for studying small systems over short time scales. In order to accurately examine the TEC and minimize the statistical errors, large systems are needed to capture atomic deflections with the temperature rise. Moreover, simulations should be conducted over long time periods to obtain smooth results with low statistical errors. These requirements make the DFT method prohibitively expensive from the computational point of view. In order to effectively address the issues of the accuracy/flexibility and expensive computational cost associated with classical MD and DFT methods in the evaluation of TEC, respectively, we propose to employ machine-learning interatomic potentials (MLIPs). As an extensive case study, herein we explore the TEC of several carbon-based monolayers at wide range of temperatures with the aid of MLIPs. The obtained results confirm that MLIPs passively fitted over inexpensive ab-initio molecular dynamics datasets can show excellent accuracy in the evaluation of complex phonon dispersion relations and also enable the examination of TEC. The proposed approach is believed to offer a simple, accurate and fast alternative of examining the TEC, highly appealing for the rapidly growing field of 2D materials.

## 2. Computational methods

This work is based on Moment Tensor Potentials (MTPs) [7,8], which represent an accurate and efficient MLIP. MTPs for carbon-based 2D systems were parameterized by fitting to the training datasets provided by ab-initio molecular dynamics (AIMD) simulations using the MLIP package [9]. AIMD calculations were carried out using the *Vienna Ab-initio Simulation Package* (VASP) [10–12] on the basis of generalized gradient approximation (GGA) with Perdew–Burke–Ernzerhof (PBE) [13] and plane-wave cutoff energy of 500 eV. In this work in order to reach maximized precision, for every 2D lattice we develop a specific MTP. This approach reduces the transferability of trained MTPs for other structures but maximizes the accuracy for a specific lattice. Worthy to note that the computational cost for training a MTP is only a fraction of subsequent classical MD simulations and like that this approach is also computationally efficient. The energy minimized unitcells by DFT calculations are replicated and taken as the input structures for AIMD calculations. To prepare the energy minimized structures with PBE/GGA functional, atomic positions and lattice sizes are altered using the conjugate gradient algorithm until Hellman-Feynman forces drop below 0.001 eV/Å using a 15×15×1 Monkhorst-Pack [14] K-point grid. The time step of AIMD calculations was set to 1 fs and a 2×2×1 Monkhorst-Pack [14] grid was employed. Along all directions, periodic boundary conditions were applied with a 14 Å vacuum layer to avoid interactions along the monolayers' normal direction. The temperature of AIMD calculations was increased gradually from 10 to 1700 K over 3000 time steps. From the complete AIMD trajectories, 500 configurations were subsampled to train the first MTPs. The accuracy of the first passively-fitted MTP was then evaluated over the complete AIMD trajectory and atomic configurations with worst extrapolation grades [15,16] were selected. Those configurations were added to the preliminary subsampled AIMD datasets and the second and final MTPs were fitted. This methodology ensures the optimal usage of full AIMD trajectories and lead to the development of more stable MTPs. Phonon dispersion relations were acquired by density functional perturbation theory calculations over supercells with a 3×3×1 Monkhorst-Pack [14] grid using the VASP and PHONOPY code[17]. Phononic properties were also examined using the parametrized MTPs and PHONOPY code, as explained in our earlier work [18].

After the development of accurate and stable MTPs, molecular dynamics simulations were carried out to explore the temperature dependent TEC of the considered nanosheets. To this end, we employed the *Large-scale Atomic/Molecular Massively Parallel Simulator* (LAMMPS) [19] package along with the developed MTPs to model interatomic interactions. Periodic boundary conditions were applied along all the directions, and the MD time step was set to 0.5 fs. The structural relaxation was conducted by employing the Nosé-Hoover thermostat and barostat method (NPT). In order to avoid undesired folding and buckling of the nanosheets during the simulations, a negligible stretching stress was adopted in the NPT setup. The temperature of the system was increased from 50 K with a fixed step of 25 K. At every temperature, NPT calculations were first conducted for 10 ps and subsequently for 50 ps. In the second NPT calculations, the sizes of the simulation box were recorded and averaged to evaluate the TEC. In order to reach smooth relations, for every structure the simulations were carried out for five times with uncorrelated initial velocities and the results were averaged. Examples of complete LAMMPS input files for the evaluation of thermal expansion of considered 2D lattices are included in the data availability section. The OVITO [20] package was employed to plot the molecular dynamics models.

## 3. Results and discussions

Our goal is to develop MLIPs for the modeling of TEC of carbon-based 2D materials using the classical MD modeling. The proposed approach can be employed for the modeling of TEC of bulk or other 2D lattices. In Fig. 1, the considered nanomembranes in this study are illustrated. We study full carbon-based allotropes of graphene, haeckelite [21], phagraphene [22], pentagraphene [23] and binary systems of $C_3N$ [24], $BC_3$, $C_2N$ [25], $C_7N_6$ [26] and hexagonal boron-nitride (h-BN) and the ternary lattice of $BC_6N$ [27]. Haeckelite and phagraphene can be considered as defective graphene structures, which are made of a combination of heptagon, hexagon and pentagon rings. Pentagraphene is a buckled and metastable hybrid $sp^2$ and $sp^3$ full carbon allotrope. It is noteworthy that carbon nitride nanosheets in recent years have been attracting remarkable attentions, and various lattices of $C_3N$ [24], $C_2N$ [25], $C_3N_{4.8}$ [28], $C_3N_3$ [29], $C_3N_4$ [30] and $C_3N_5$ [31] have been experimentally realized. Among the aforementioned nanomembranes, only the $C_3N$ shows graphene-like structure and the rest include nanoporous lattices [32]. Therefore, the analysis of TEC of $C_2N$ [25] and $C_7N_6$ [26] monolayers with nanoporous structures can provide useful insights. The hexagonal lattice constants of graphene, h-BN, haeckelite, $C_2N$, $C_3N$, $BC_3$, $C_7N_6$ and $MoS_2$ are found to be 2.47, 2.51, 7.09, 8.33, 4.86, 5.17 and 6.79 Å, respectively. The rectangular unit cell parameters for phagraphene are 8.11 and 6.64 Å and for $BC_6N$ are 2.47 and 8.64 Å and for pentagraphene is 3.64 Å.

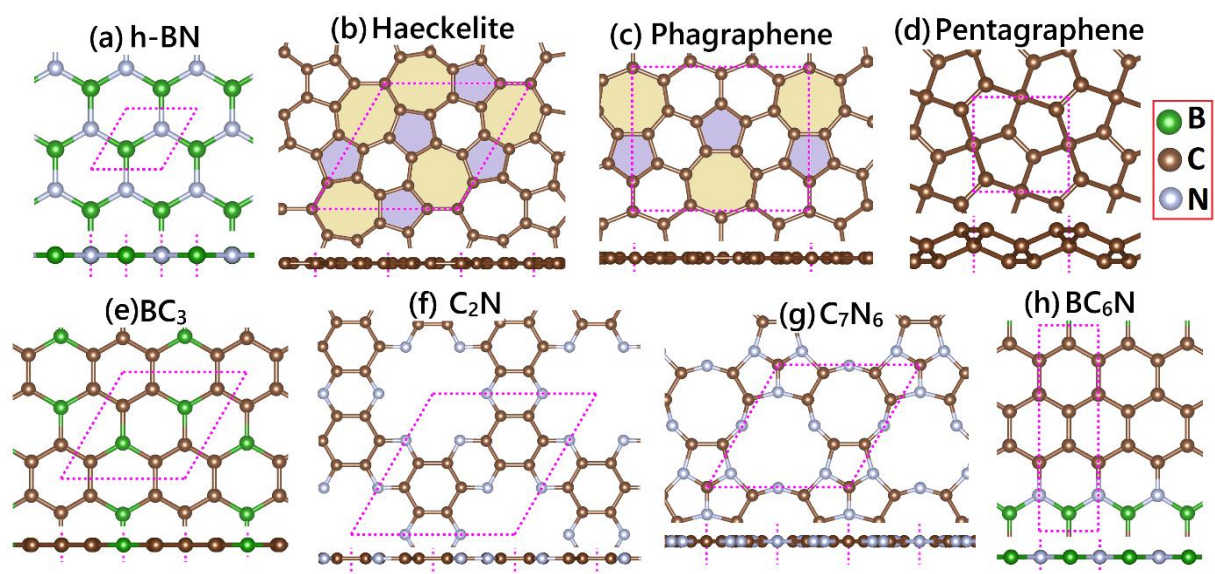

**Fig. 1**, Top and side views of the considered monolayers in this work; (a) h-BN or graphene (b) haeckelite[21] (c) phagraphene[22] (d) pentagraphene [23] (e) $C_3N$ [24] or $BC_3$ (f) $C_2N$ [25] (g) $C_7N_6$ [26] and (h) $BC_6N$ [27]. The dashed lines depict the atomic unit cell.

As mentioned earlier, the training datasets were prepared using the AIMD simulations over supercell systems. For graphene, h-BN, phagraphene, haeckelite, pentagraphene, $C_2N$, $C_3N$ and $BC_3$, we used $n \times n \times 1$ supercell, with $n$ equal to 6, 6, 2, 2, 3, 2, 3 and 3, respectively. For the $C_7N_6$ and $BC_6N$ monolayers, $2 \times 3 \times 1$ and $5 \times 2 \times 1$ supercells were modeled, respectively. In general, for carbon based nanosheets the size of the simulation box along each direction was considered to be over 10 Å. In the preparation of dataset with AIMD calculations the inclusion of larger supercells can enhance the accuracy and transferability of the trained MTPs, because they can better capture large deflection, but this reduces the computational efficiency of the training procedure. MTPs with the fixed cutoff distance of 3.5 Å and 125, 221 and 381 parameters were fitted for full-carbon, binary and $BC_6N$ systems, respectively. In order to evaluate the accuracy of fitted MTPs, in Fig. 2 the MTP-based phonon dispersion relations are compared with those acquired by DFT for the considered monolayers. To obtain phonon dispersions with DFT, supercell sizes of $n \times n \times 1$, with $n$ equal to 8, 8, 3, 3, 5, 3, 4, 4 and 3, for graphene, h-BN, phagraphene, haeckelite, pentagraphene, $C_2N$, $C_3N$, $BC_3$ and $C_7N_6$, respectively, were employed. For the $BC_6N$ monolayer an $8 \times 2 \times 1$ supercell was considered. From the results shown in Fig. 2, close agreement between MTP- and DFT-based results are observable, particularly for the acoustic modes which are the dominant heat carriers in these systems. In particular, the dispersion of the ZA acoustic mode are accurately reproduced by the trained MTPs. Despite the inclusion of rather small supercells in the preparation of dataset, these results reveal that out-of-plane and long-wave length vibrations can be described accurately with trained MTPs. By considering the results for haeckelite and $BC_6N$, it is also noticeable that the accuracy of developed models is independent of the number of element in the systems. All trained potentials in MTP native format are included in the data availability section of this manuscript. The presented results highlight the remarkable accuracy of the trained MTPs and confirm their validity for the evaluation of TEC. After ensuring the accuracy and stability of the trained MTPs, we now investigate the TEC ($\alpha$) using the following equation:

$$\alpha = \frac{1}{A}\frac{dA}{dT} \qquad (1)$$

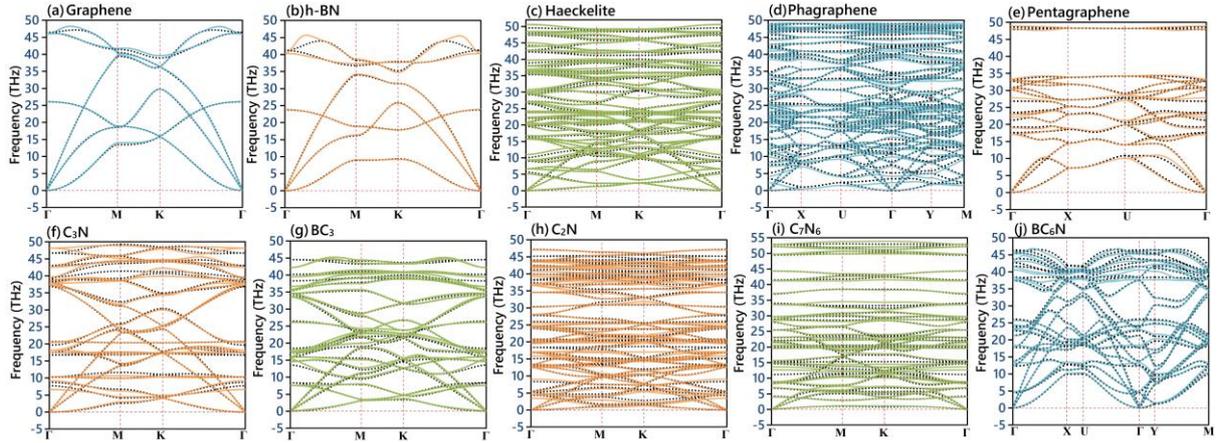

**Fig. 2**, Phonon dispersion relations predicted by MTP (black dotted lines) and DFT (continuous lines).

wherein *A* denotes the projected area of the monolayers and *T* is the temperature. In order to calculate the TEC, we plot the area of each monolayer at various temperatures, starting from 50 K to 1000 K with a fixed 25 K step. The projected area is simply calculated using the in-plane dimensions of the simulation box. By calculating the slope of the measured area as a function of temperature, the TEC curve can be acquired based on the Eq. 1. Fig. 3 shows examples of the atomic configurations of four different monolayers after the structural equilibration at various temperatures. As can be clearly seen, as a result of thermal effects and contact with vacuum along the monolayers normal direction, wrinkles form throughout the structures which subsequently result in the decrease of projected area in comparison with fully flat structures at 0 K. As shown for the case of pentagraphene, by increasing the temperature the nanosheets may also disintegrate.

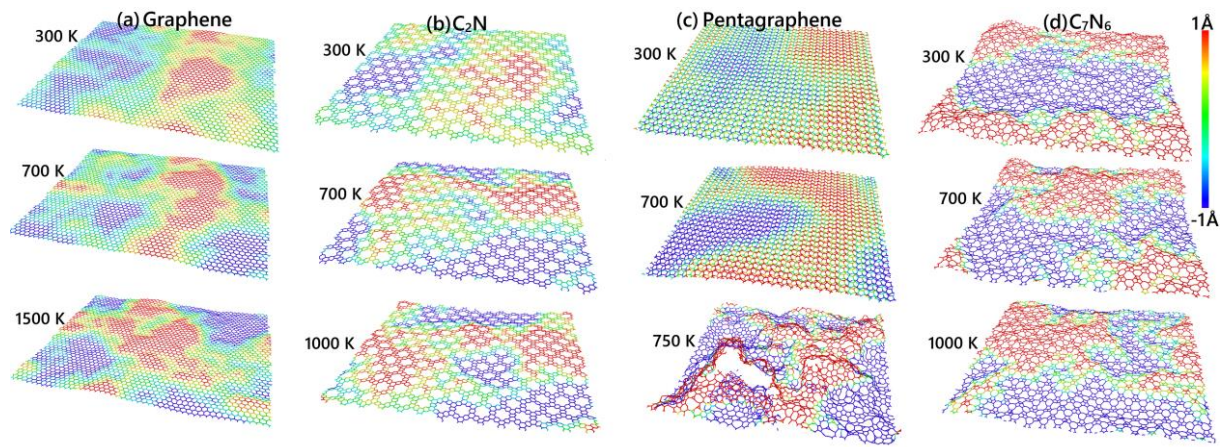

**Fig. 3**, Atomic configurations of single-layer (a) graphene, (b) $C_2N$, (c) pentagraphene and (d) $C_7N_6$ at various temperatures. Color coding represents the out of plane displacement.

First, we compare our estimation for the graphene TEC at 300 K with the available studies in the literature, as summarized in Table 1. It is observable that MTP-based prediction for TEC of graphene, $-2.95\times10^{-6}$ K$^{-1}$, is close to the mean value of DFT-based reports of $-3.6\times10^{-6}$ [2] and $-2.95\times10^{-6}$ K$^{-1}$ [16] and is distinctly higher than that predicted by MD with AIREBO empirical potential. It is also noticeable that our prediction is very close to the recent study [16] on the basis of Gaussian approximation potential (GAP). In comparison with the earlier MLIP-based work with GAP [16], the trained MTP in this work could slightly better reproduce the phonon dispersions of acoustic modes in graphene, particularly the out-of-plane ZA mode. These out-of-plane vibrations are the main source of negative TEC in 2D systems. On the other side, there exists a considerable gap between the theoretical predictions and the experimental measurement by Yoon *et al.* [3], which as discussed in a recent electron diffraction study [6] is mainly attributed to the effect of substrate. We remind that in the conducted theoretical works the pristine and suspended monolayers have been considered and such that effects of point defects and grain boundaries, substrate and also the contaminations commonly existing in fabricated graphene nanosheets have been neglected. As discussed earlier, phagraphene and haeckelite resemble, from the structural points of view, a highly defective graphene, and for these monolayers at 300 K we found the close values of TECs, around $-6.5\times10^{-6}$ K$^{-1}$, which is considerably closer to the experimental findings.

Table 1: Comparison of the predicted thermal expansion coefficient of graphene at 300 K.

| Study | $\alpha$ (K$^{-1}$) at 300 K |
|---|---|
| Present work (MD with MTP) | $-2.95\times10^{-6}$ |
| Mounet et al. (DFT) [2] | $-3.6\times10^{-6}$ |
| Ghasemi *et al.* (MD with AIREBO potential) [33] | $-2.10\times10^{-6}$ |
| Yoon et al. (Experiment) [3] | $-8.0\times10^{-6}$ |
| Demiroğlu et al. (DFT) [16] | $-2.46\times10^{-6}$ |
| Demiroğlu et al. (MD with GAP)[16] | $-2.83\times10^{-6}$ |

In Fig. 4 and 5 we plot the projected area variation of the considered monolayers normalized by the number of atoms as a function of temperature. It is obvious that the projected area of the majority of two-dimensional structures decreases with increasing temperature, due to the general increase of the wrinkles formation with the temperature rise up to 1000 K. Nonetheless, with increasing temperature, the real surface area of each structure increases due to higher atomic vibrations. While the majority of the considered systems show the overall decreasing trend of projected area by increasing the temperature, as observable from the results in Fig. 5, $C_7N_6$ and pentagraphene monolayers exhibit different behavior. For the case of pentagraphene, the overall increase of the projected area with increasing temperature is insignificant, which implies low TEC. In this system as shown in Fig. 3C, because of the 3D-like lattice, the structure shows high bending rigidity, which limits the formation of the out-of-plane wrinkles by increasing the temperature. On the other side, $C_7N_6$ monolayer exhibits very different mechanism than pentagraphene. In $C_7N_6$ nanosheet, due to the existence of nanoporosity in the lattice and irregular and non-hexagonal bonding configurations, the out-of-plane acoustic mode appears with a very flat dispersion and the structure shows considerable wrinkling at low temperature (see Fig. 3a). In this monolayer, with the increase in temperature due to the interlayer van der Waals interactions noticeably the wrinkling is suppressed, which results in the extension of the projected area.

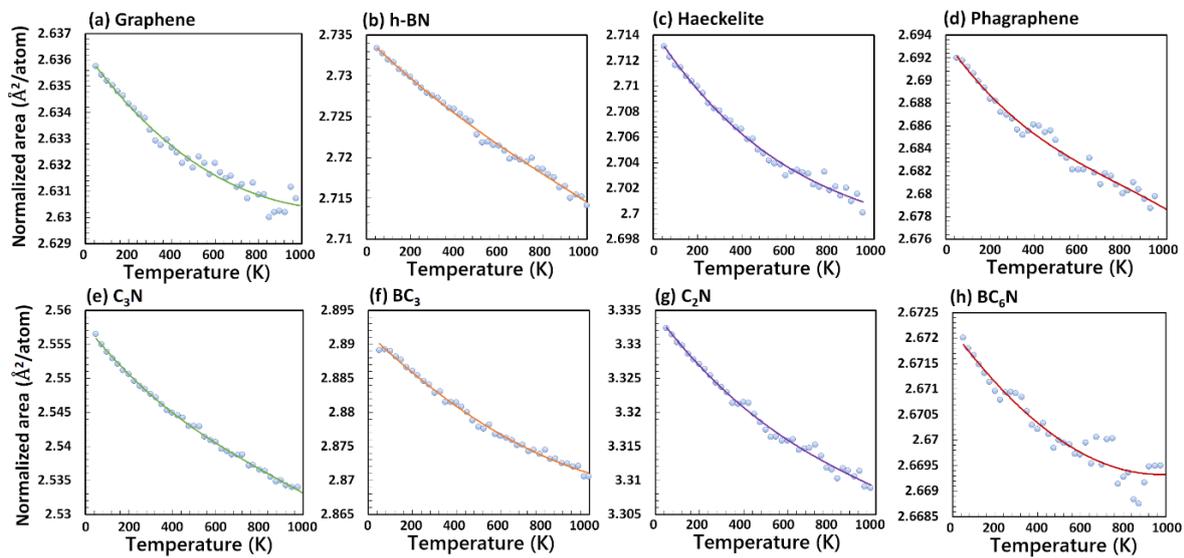

Fig. 4, In graphene, h-BN, haeckelite, phagraphene, $C_3N$, $BC_3$, $C_2$ and $BC_6N$ monolayers the projected area of the system decreases with the increase of temperature.

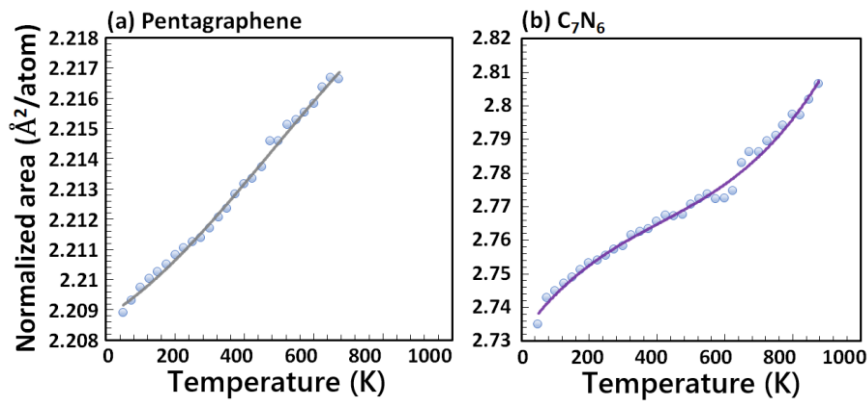

Fig. 5, In (a) pentagraphene and (b) $C_7N_6$ the projected area of the system increases with the increase of temperature.

Fig. 6 and Table 2 summarize the results of the predicted TEC values for considered 2D structures as a function of temperature. This illustration contains two parts for the structures with (a) negative and (b) positive thermal expansion. As can be seen, the thermal expansion coefficient for the majority of structures is negative and ranges around $1\times10^{-6}$ K$^{-1}$. Differences in TEC of various structures can be attributed to the atomic types, atomic configurations, bond strengths, wrinkle magnitude and density. The results shown in Fig. 6a also reveal that the TEC at high temperatures tends to get closer to zero and shows a relatively plateau pattern. This is due to the fact that at high temperatures the production of wrinkles in monolayers tend to reach a saturation level due to the interlayer van der Waals interactions with neighboring wrinkles. As such by increasing the temperature, generally, the TEC first reaches a plateau and later may increases because the real surface increases with the temperature. As shown in Fig. 6a, below 1000 K such a process is easily observable for the cases of phagraphene, h-BN, $C_3N$, $C_2N$ monolayers. As shown in Fig. 6b, for the case of pentagraphene due to 3D-structure and remarkable out-of-plane rigidity, TEC is positive but remains close to zero. Moreover, for the $C_7N_6$ monolayer, due to the formation of remarkable wrinkles at low temperature, the wrinkling saturation occurs at low temperatures and subsequently yields a positive TEC curve. Results summarized in Table 2 clearly show higher negative TEC values in h-BN, Haeckelite, phagraphene, $C_3N$, $BC_3$ and $C_2N$ monolayers than graphene. It is noticeable that $BC_6N$ monolayer which in fact shows two rings wide zigzag-edge graphene nanoribbons connected by polyacetylene-like BN chains, exhibits the closet behavior to graphene.

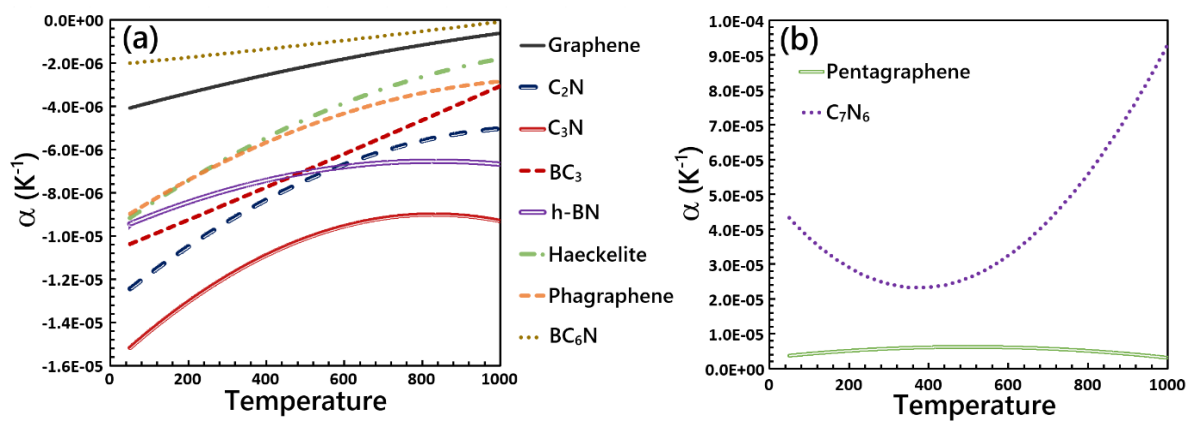

**Fig. 6**, Thermal expansion coefficient as a function of temperature for monolayers with (a) negative α values, graphene, h-BN, Haeckelite, phagraphene, $C_3N$, $BC_3$, $C_2N$ and $BC_6N$ and (b) positive α values, $C_7N_6$ and pentagraphene.

Table 2: Comparison of the predicted thermal expansion coefficients of considered monolayers at different temperatures.

| System | $\alpha$ ($10^{-6}$ K$^{-1}$) | | | | |
|---|---|---|---|---|---|
| | 200 K | 300 K | 400 K | 500 K | 1000 K |
| Graphene | -2.39 | -2.95 | -2.55 | -2.16 | -0.6 |
| C$_2$N | -10.50 | -9.36 | -8.30 | -7.47 | -5.03 |
| C$_3$N | -13.00 | -11.90 | -10.90 | -10.10 | -9.31 |
| BC$_3$ | -9.26 | -8.51 | -7.75 | -6.98 | -3.06 |
| h-BN | -8.44 | -7.87 | -7.41 | -7.06 | -6.68 |
| Haeckelite | -7.45 | -6.41 | -5.46 | -4.60 | -1.80 |
| Phagraphene | -7.40 | -6.49 | -5.68 | -4.96 | -2.85 |
| BC$_6$N | -1.73 | -1.55 | -1.36 | -1.16 | -0.08 |
| Pentagraphene | 5.11 | 5.73 | 6.09 | 6.20 | 3.10 |
| C$_7$N$_6$ | 28.80 | 24.30 | 23.30 | 26.00 | 92.90 |

As discussed, it is clear that complex thermal expansion behavior of 2D materials is mainly attributed to the out-of-plane vibrations and thermal effects that results in local wrinkles and deflection. In order to take into account this effect, one approach is to calculate the real surface area of the nanosheets in the evaluation of TEC. To this end, we conducted curve fitting to the atomic positions on the basis of nearest neighbor approximation to measure the real surface area as shown in Fig. 7a. In Fig. 7b to 7e, we plot the evolution of the real area as a function of temperature for graphene, BC$_3$, h-BN and C$_3$N monolayers. In contrast with the results presented in Fig. 5, the real surface area shows a clear increase with the increase in the temperature. Insets of Fig. 7 show the thermal expansion curves of the considered monolayers that reveal positive values for TEC. These results once again confirm the dominant role of the out-of-plane deflections that results in complex thermal expansion behavior of 2D materials, otherwise the TEC of 2D materials would generally be positive like that of the conventional bulk materials, provided that the real surface area is considered in the evaluation. With the corrected surface area, the predicted TEC of graphene, BC$_3$, h-BN, C$_2$N and C$_3$N monolayers at room temperature all become positive and 400, 315, 365, 447 and 420×10$^{-6}$ K$^{-1}$, respectively, which unlike our earlier estimations are remarkably. Apart from this observation, our study highlights that MLIPs offer computationally efficient and versatile possibility to explore thermal expansion of complex nanomaterials.

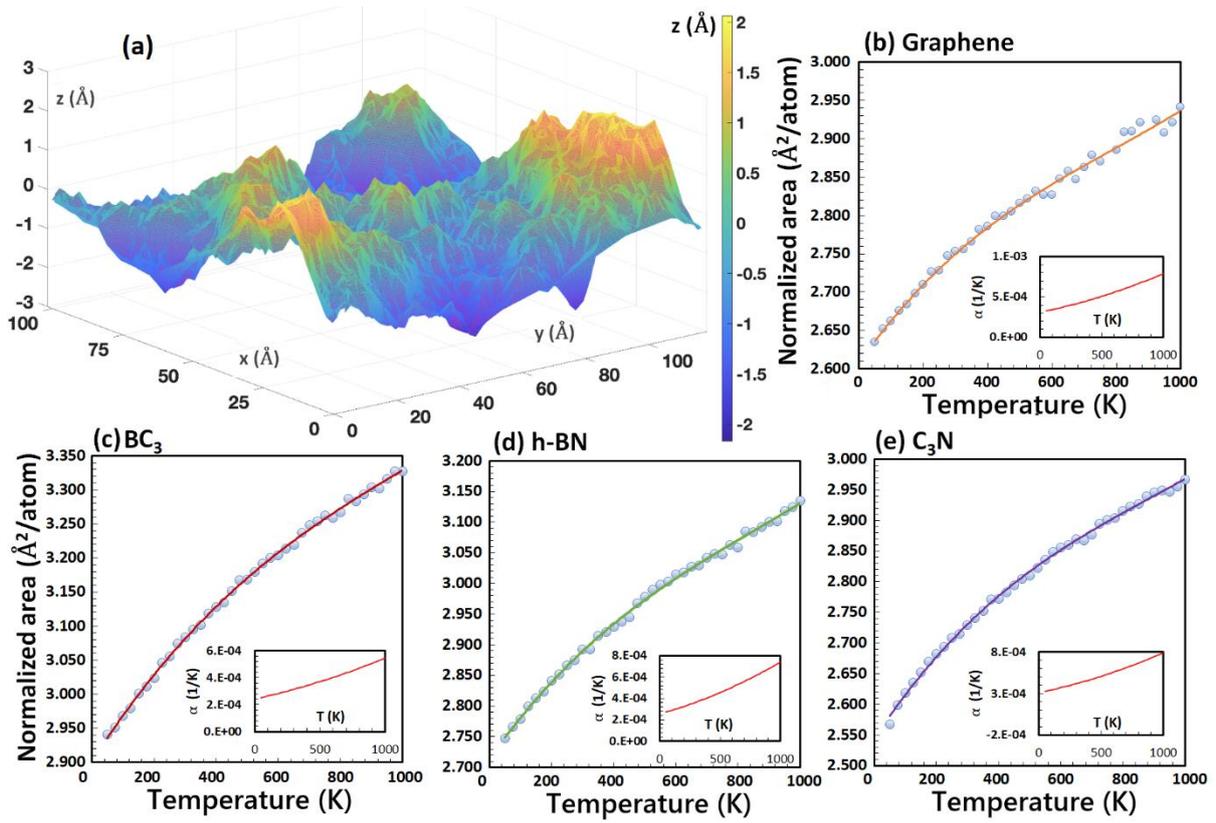

Fig. 7, (a) Curve fitting to calculate the real surface area of monolayers. Evolution of real area as a function of temperature for (b) graphene, (c) BC$_3$, (d) h-BN and (e) C$_3$N monolayers. Inset shows the calculated thermal expansion coefficient $\alpha$ as a function of temperature.

## 4. Conclusion

Our study confirms that moment tensor potentials passively fitted over computationally inexpensive ab-initio molecular dynamics dataset enable the accurate examination of phonon dispersion relations and thermal expansion coefficient at a wide range of temperatures. We show that with the aid of the developed interatomic potentials, the thermal effects and complex atomic vibration can be accurately explored with the standard classical molecular dynamics approach. Our analysis also confirms that the accuracy of developed models is convincingly independent of the number of constituent elements and complexity of atomic lattice. It is thus believed that the proposed approach can be considered as an efficient and accurate methodology to examine the thermal expansion of various structures and in particular two-dimensional materials.

## Acknowledgment

B.M. and X.Z. appreciate the funding by the Deutsche Forschungsgemeinschaft (DFG, German Research Foundation) under Germany's Excellence Strategy within the Cluster of Excellence

PhoenixD (EXC 2122, Project ID 390833453). A.V.S. was supported by the Russian Science Foundation (Grant No 18-13-00479, https://rscf.ru/project/18-13-00479/). B. M. is also especially thankful to the VEGAS cluster at Bauhaus University of Weimar and Dr. Chernenko for the support of this study. A. R. appreciates Imam Khomeini International University Research Council for support of this study. Authors also acknowledge the support of the cluster team at the Leibniz Universität of Hannover.

### Data availability

All trained MTPs and examples of LAMMPS input files for the evaluation of thermal expansion are accessible via the following Mendeley dataset: http://dx.doi.org/10.17632/4ftjdf93v7.1

### References

[1] K.S. Novoselov, A.K. Geim, S. V Morozov, D. Jiang, Y. Zhang, S. V Dubonos, I. V Grigorieva, A.A. Firsov, Electric field effect in atomically thin carbon films., Science. 306 (2004) 666–9. https://doi.org/10.1126/science.1102896.

[2] N. Mounet, N. Marzari, First-principles determination of the structural, vibrational and thermodynamic properties of diamond, graphite, and derivatives, Phys. Rev. B. 71 (2005) 205214. https://doi.org/10.1103/PhysRevB.71.205214.

[3] D. Yoon, Y.-W. Son, H. Cheong, Negative Thermal Expansion Coefficient of Graphene Measured by Raman Spectroscopy, Nano Lett. 11 (2011) 3227–3231. https://doi.org/10.1021/nl201488g.

[4] S. Linas, Y. Magnin, B. Poinsot, O. Boisron, G.D. Förster, V. Martinez, R. Fulcrand, F. Tournus, V. Dupuis, F. Rabilloud, L. Bardotti, Z. Han, D. Kalita, V. Bouchiat, F. Calvo, Interplay between Raman shift and thermal expansion in graphene: Temperature-dependent measurements and analysis of substrate corrections, Phys. Rev. B. 91 (2015) 75426. https://doi.org/10.1103/PhysRevB.91.075426.

[5] S. Tian, Y. Yang, Z. Liu, C. Wang, R. Pan, C. Gu, J. Li, Temperature-dependent Raman investigation on suspended graphene: Contribution from thermal expansion coefficient mismatch between graphene and substrate, Carbon N. Y. 104 (2016) 27–32. https://doi.org/https://doi.org/10.1016/j.carbon.2016.03.046.

[6] E. Kano, M. Malac, M. Hayashida, Substrate and contamination effects on the thermal expansion coefficient of suspended graphene measured by electron diffraction, Carbon N. Y. 163 (2020) 324–332. https://doi.org/https://doi.org/10.1016/j.carbon.2020.02.008.

[7] A. V. Shapeev, Moment tensor potentials: A class of systematically improvable interatomic potentials, Multiscale Model. Simul. 14 (2016) 1153–1173. https://doi.org/10.1137/15M1054183.

[8] K. Gubaev, E. V. Podryabinkin, G.L.W. Hart, A. V. Shapeev, Accelerating high-throughput searches for new alloys with active learning of interatomic potentials, Comput. Mater. Sci. 156 (2019) 148–156. https://doi.org/10.1016/j.commatsci.2018.09.031.

[9] A.S. Ivan Novikov, Konstantin Gubaev, Evgeny Podryabinkin, The MLIP package: Moment Tensor Potentials with MPI and Active Learning, Mach. Learn. Sci. Technol. 2

(2021) 025002. http://iopscience.iop.org/article/10.1088/2632-2153/abc9fe.

[10]  G. Kresse, J. Furthmüller, Efficiency of ab-initio total energy calculations for metals and semiconductors using a plane-wave basis set, Comput. Mater. Sci. 6 (1996) 15–50. https://doi.org/10.1016/0927-0256(96)00008-0.

[11]  G. Kresse, J. Furthmüller, Efficient iterative schemes for ab initio total-energy calculations using a plane-wave basis set, Phys. Rev. B. 54 (1996) 11169–11186. https://doi.org/10.1103/PhysRevB.54.11169.

[12]  G. Kresse, From ultrasoft pseudopotentials to the projector augmented-wave method, Phys. Rev. B. 59 (1999) 1758–1775. https://doi.org/10.1103/PhysRevB.59.1758.

[13]  J. Perdew, K. Burke, M. Ernzerhof, Generalized Gradient Approximation Made Simple., Phys. Rev. Lett. 77 (1996) 3865–3868. https://doi.org/10.1103/PhysRevLett.77.3865.

[14]  H. Monkhorst, J. Pack, Special points for Brillouin zone integrations, Phys. Rev. B. 13 (1976) 5188–5192. https://doi.org/10.1103/PhysRevB.13.5188.

[15]  E. V Podryabinkin, A. V Shapeev, Active learning of linearly parametrized interatomic potentials, Comput. Mater. Sci. 140 (2017) 171–180. https://doi.org/10.1016/j.commatsci.2017.08.031.

[16]  İ. Demiroğlu, Y. Karaaslan, T. Kocabaş, M. Keçeli, Á. Vázquez-Mayagoitia, C. Sevik, Computation of the Thermal Expansion Coefficient of Graphene with Gaussian Approximation Potentials, J. Phys. Chem. C. 125 (2021) 14409–14415. https://doi.org/10.1021/acs.jpcc.1c01888.

[17]  A. Togo, I. Tanaka, First principles phonon calculations in materials science, Scr. Mater. 108 (2015) 1–5. https://doi.org/10.1016/j.scriptamat.2015.07.021.

[18]  B. Mortazavi, I.S. Novikov, E. V Podryabinkin, S. Roche, T. Rabczuk, A. V Shapeev, X. Zhuang, Exploring phononic properties of two-dimensional materials using machine learning interatomic potentials, Appl. Mater. Today. 20 (2020) 100685. https://doi.org/10.1016/j.apmt.2020.100685.

[19]  S. Plimpton, Fast Parallel Algorithms for Short-Range Molecular Dynamics, J. Comput. Phys. 117 (1995) 1–19. https://doi.org/10.1006/jcph.1995.1039.

[20]  A. Stukowski, Visualization and analysis of atomistic simulation data with OVITO–the Open Visualization Tool, Model. Simul. Mater. Sci. Eng. 18 (2009) 015012. https://doi.org/10.1088/0965-0393/18/1/015012.

[21]  H. Terrones, M. Terrones, E. Hernández, N. Grobert, J.-C. Charlier, P.M. Ajayan, New Metallic Allotropes of Planar and Tubular Carbon, Phys. Rev. Lett. 84 (2000) 1716–1719. https://doi.org/10.1103/PhysRevLett.84.1716.

[22]  Z. Wang, X.F. Zhou, X. Zhang, Q. Zhu, H. Dong, M. Zhao, A.R. Oganov, Phagraphene: A Low-Energy Graphene Allotrope Composed of 5-6-7 Carbon Rings with Distorted Dirac Cones, Nano Lett. 15 (2015) 6182–6186. https://doi.org/10.1021/acs.nanolett.5b02512.

[23]  S. Zhang, J. Zhou, Q. Wang, X. Chen, Y. Kawazoe, P. Jena, Penta-graphene: A new carbon allotrope, Proc. Natl. Acad. Sci. U. S. A. (2015). https://doi.org/10.1073/pnas.1416591112.

[24]  J. Mahmood, E.K. Lee, M. Jung, D. Shin, H.-J. Choi, J.-M. Seo, S.-M. Jung, D. Kim, F. Li, M.S. Lah, N. Park, H.-J. Shin, J.H. Oh, J.-B. Baek, Two-dimensional polyaniline ($C_3N$) from carbonized organic single crystals in solid state, Proc. Natl. Acad. Sci. . 113 (2016) 7414–7419. https://doi.org/10.1073/pnas.1605318113.

[25]  J. Mahmood, E.K. Lee, M. Jung, D. Shin, I.Y. Jeon, S.M. Jung, H.J. Choi, J.M. Seo, S.Y. Bae, S.D. Sohn, N. Park, J.H. Oh, H.J. Shin, J.B. Baek, Nitrogenated holey two-

dimensional structures, Nat. Commun. 6 (2015) 1–7. https://doi.org/10.1038/ncomms7486.

[26] B. Mortazavi, M. Shahrokhi, A. V Shapeev, T. Rabczuk, X. Zhuang, Prediction of C7N6 and C9N4: stable and strong porous carbon-nitride nanosheets with attractive electronic and optical properties, J. Mater. Chem. C. 7 (2019) 10908–10917. https://doi.org/10.1039/C9TC03513C.

[27] B. Mortazavi, Ultrahigh thermal conductivity and strength in direct-gap semiconducting graphene-like BC6N: A first-principles and classical investigation, Carbon N. Y. 182 (2021) 373–383. https://doi.org/https://doi.org/10.1016/j.carbon.2021.06.038.

[28] I.Y. Kim, S. Kim, X. Jin, S. Premkumar, G. Chandra, N.-S. Lee, G.P. Mane, S.-J. Hwang, S. Umapathy, A. Vinu, Ordered Mesoporous C3N5 with a Combined Triazole and Triazine Framework and Its Graphene Hybrids for the Oxygen Reduction Reaction (ORR), Angew. Chemie. 130 (2018) 17381–17386. https://doi.org/10.1002/ange.201811061.

[29] J. Zeng, Z. Chen, X. Zhao, W. Yu, S. Wu, J. Lu, K.P. Loh, J. Wu, From All-Triazine C3N3 Framework to Nitrogen-Doped Carbon Nanotubes: Efficient and Durable Trifunctional Electrocatalysts, ACS Appl. Nano Mater. 2 (2019) 12. https://doi.org/10.1021/acsanm.9b02011.

[30] L.F. Villalobos, M.T. Vahdat, M. Dakhchoune, Z. Nadizadeh, M. Mensi, E. Oveisi, D. Campi, N. Marzari, K.V. Agrawal, Large-scale synthesis of crystalline g-C3N4 nanosheets and high-temperature H2 sieving from assembled films, Sci. Adv. 6 (2020) eaay9851. https://doi.org/10.1126/sciadv.aay9851.

[31] P. Kumar, E. Vahidzadeh, U.K. Thakur, P. Kar, K.M. Alam, A. Goswami, N. Mahdi, K. Cui, G.M. Bernard, V.K. Michaelis, K. Shankar, C3N5: A Low Bandgap Semiconductor Containing an Azo-Linked Carbon Nitride Framework for Photocatalytic, Photovoltaic and Adsorbent Applications, J. Am. Chem. Soc. 141 (2019) 5415–5436. https://doi.org/10.1021/jacs.9b00144.

[32] S.M. Hatam-Lee, A. Rajabpour, S. Volz, Thermal conductivity of graphene polymorphs and compounds: From C3N to graphdiyne lattices, Carbon N. Y. 161 (2020) 816–826. https://doi.org/10.1016/j.carbon.2020.02.007.

[33] H. Ghasemi, A. Rajabpour, A novel approach to calculate thermal expansion of graphene: Molecular dynamics study, Eur. Phys. J. Plus. 132 (2017) 221. https://doi.org/10.1140/epjp/i2017-11491-y.